\documentclass[english]{article}
\usepackage[T1]{fontenc}
\usepackage[latin9]{inputenc}
\usepackage{amsmath}
\usepackage{amssymb}
\usepackage{babel}

\begin{document}

\title{Similarities between neutrinos and Higgs bosons within a chiral supersymmetric
formulation}

\author{Tobias Gleim}
\maketitle
\begin{abstract}
Taking into account the helicity of a massless particle, which obeys
a Dirac equation and is exposed to an electromagnetic field, one soon
arrives at a Lagrangian containing a chiral supersymmetric operator.
We can even achieve an analogous result in case of an electroweak
interaction. Then the Lagrangian contains terms that look very similar
to those usually being interpreted as mass terms for the \emph{W}
and \emph{Z} bosons appearing in models of spontaneous symmetry breaking,
but this time they are accompanied by neutrino fields instead of Higgs
fields. This invites us to a speculation that in the procedure of
spontaneous symmetry breaking, neutrinos could take on the role, which
we normally ascribe to the Higgs bosons. 
\end{abstract}
With the intention to reformulate the theory of electroweak interaction
as described within the frame of the Standard Model, we would like
to start from a Dirac equation extended to fermion multiplets:

\begin{equation}
\psi=\left(\begin{array}{c}
\psi_{1}\\
...\\
\psi_{N}\end{array}\right),\label{eq:1}\end{equation}
i.e. a wave operator whose components $\psi_{1}$, ..., $\psi_{N}$
themselves are 4-spinors and thus obey a Dirac equation. Now, we consider
a coupling to a not necessarily Abelian field with potential $M_{\mu}$:

\begin{equation}
\left[\gamma^{\mu}\left(\hat{p}_{\mu}+M_{\mu}\right)-m\right]\psi=0,\label{eq:2}\end{equation}
where $\hat{p}_{\mu}=i\partial_{\mu}$ and $m$ is a (in general matrix-like
and assumed to be diagonal and real) constant. \eqref{eq:2} can be
used for the case of an electron in an electromagnetic field with
potential $A_{\mu}$: then $\psi$ has only one component, $m$ is
the electron mass and $M_{\mu}=A_{\mu}$ is a scalar. But \eqref{eq:2}
could also be applied in case of an electroweak interaction: then
$M_{\mu}$ belongs to a non-Abelian field (and $m$ is set to zero)
and $\psi$ contains three components:

\begin{equation}
\psi=\left(\begin{array}{c}
\nu_{L}\\
e_{L}\\
e_{R}\end{array}\right),\label{eq:3}\end{equation}
where the first two designate a doublet of left-handed leptons, i.e.
left-handed neutrino and electron wave operators, $\nu_{L}$ and $e_{L}$
respectively, whereas the last one belongs to a singlet containing
a right-handed electron with a wave operator $e_{R}$ (here and in
the following we restrict ourselves to the first generation of leptons
and omit the quarks for the sake of simplicity). Since the potential
$M_{\mu}$ is constructed in a way that there is only a coupling between
the left-handed leptons of the doublet, but no coupling between the
left-handed doublet and the right-handed singlet, it is adequate to
inspect left- and right-handed spinors independently from each other.
Hence we use

\begin{equation}
\psi_{\pm}=P_{\pm}\psi,\label{eq:4}\end{equation}
which define the right- and left-handed projections of $\psi$ with
the help of the projection operators

\begin{equation}
P_{\pm}=\frac{1}{2}\left(1\pm\gamma^{5}\right),\label{eq:5}\end{equation}
i.e. $P_{+}+P_{-}=1$, where 

\begin{equation}
\gamma^{5}=i\gamma^{0}\gamma^{1}\gamma^{2}\gamma^{3}\label{eq:6}\end{equation}
with the properties

\begin{equation}
\left[\gamma^{\mu},\gamma^{5}\right]_{+}=0,\;\left(\gamma^{5}\right)^{2}=1,\;\left(\gamma^{5}\right)^{\dagger}=\gamma^{5}.\label{eq:7}\end{equation}
$\left[\cdot,\cdot\right]_{+}$ designates the anti-commutator, whereas
$\left[\cdot,\cdot\right]_{-}$ is the commutator. Due to \eqref{eq:7}, 

\begin{equation}
P_{\pm}^{\dagger}=P_{\pm},\; P_{\pm}P_{\mp}=0,\; P_{\pm}^{2}=P_{\pm},\label{eq:8}\end{equation}
as well as the following relations hold:

\begin{equation}
P_{\pm}\gamma^{\mu}=\gamma^{\mu}P_{\mp}.\label{eq:9}\end{equation}

Now, it is our aim to find an operator equation for $\psi_{\pm}$
that resembles the one for Higgs bosons, which is Klein-Gordon-like.
A Klein-Gordon-like equation for $\psi_{\pm}$ can be obtained in
the subsequent way (see e.g. \cite{key-1}):

\begin{equation}
m\psi=m\left(P_{\pm}+P_{\mp}\right)\psi=\left[\gamma^{\mu}\left(\hat{p}_{\mu}+M_{\mu}\right)+m\right]P_{\pm}\psi=0,\label{eq:10}\end{equation}
where we have used \eqref{eq:2} and \eqref{eq:9}. Applying the Dirac-operator
to \eqref{eq:10}, we can regard the following equation

\begin{equation}
0=m\left[\gamma^{\mu}\left(\hat{p}_{\mu}+M_{\mu}\right)-m\right]\psi=\left[\gamma^{\mu}\gamma^{\nu}\left(\hat{p}_{\mu}+M_{\mu}\right)\left(\hat{p}_{\nu}+M_{\nu}\right)-m^{2}\right]P_{\pm}\psi\label{eq:11}\end{equation}
as the one we have been searching for, if we use

\begin{equation}
\gamma^{\mu}\gamma^{\nu}=g^{\mu\nu}-i\sigma^{\mu\nu},\label{eq:13a}\end{equation}
because of

\begin{equation}
\left[\gamma^{\mu},\gamma^{\nu}\right]_{+}=2g^{\mu\nu}\label{eq:12}\end{equation}
and

\begin{equation}
\sigma^{\mu\nu}=\frac{i}{2}\left[\gamma^{\mu},\gamma^{\nu}\right]_{-}=-\sigma^{\nu\mu},\label{eq:13b}\end{equation}
which yields:

\begin{equation}
0=\left[\left(\hat{p}_{\mu}+M_{\mu}\right)\left(\hat{p}^{\mu}+M^{\mu}\right)-m^{2}+\frac{1}{2}\sigma^{\mu\nu}E_{\mu\nu}\right]P_{\pm}\psi\label{eq:14a}\end{equation}
or

\begin{equation}
0=\left[\hat{p}_{\mu}\hat{p}^{\mu}-m^{2}+\left(\hat{p}_{\mu}M^{\mu}\right)+2M_{\mu}\hat{p}^{\mu}+M_{\mu}M^{\mu}+\frac{1}{2}\sigma^{\mu\nu}E_{\mu\nu}\right]P_{\pm}\psi,\label{eq:14b}\end{equation}
where the operator $\hat{p}_{\mu}$in the third term should only act
on $M^{\mu}$, but not on $\psi$, and

\begin{equation}
E_{\mu\nu}=\left(\partial_{\mu}M_{\nu}-\partial_{\nu}M_{\mu}\right)-i\left[M_{\mu},M_{\nu}\right]_{-}.\label{eq:14c}\end{equation}
The adjoint equation that belongs to \eqref{eq:14a} can be obtained
by taking the hermitian conjugate of \eqref{eq:14a}, which results
in

\begin{equation}
0=\left[\left(\hat{p}^{\mu}-M^{\mu T}\right)\left(\hat{p}_{\mu}-M_{\mu}^{T}\right)-m^{2}\right]\bar{\psi}_{\pm}+\frac{1}{2}\bar{\psi}_{\pm}\sigma^{\mu\nu}E_{\mu\nu},\label{eq:15}\end{equation}
where $M_{\mu}^{T}$ is the transposed matrix of $M_{\mu}$,

\begin{equation}
\bar{\psi}_{\pm}=\psi_{\pm}^{\dagger}\gamma^{0}=\bar{\psi}P_{\mp}\label{eq:16}\end{equation}
and where we have applied

\begin{equation}
\gamma_{\mu}^{\dagger}=\gamma^{0}\gamma^{\mu}\gamma^{0}\label{eq:17}\end{equation}
as well as \eqref{eq:8}, \eqref{eq:9} and \eqref{eq:12}.

\section*{Connection with chiral supersymmetry}

Operators like the ones in \eqref{eq:10} and \eqref{eq:11} are well-known
from chiral supersymmetric formulations, if \emph{$m$} is set to
zero and $M_{\mu}$ is e.g. identified with an electromagnetic potential
$M_{\mu}$. There (see e.g. \cite{key-2}), two operators are defined:

\begin{equation}
\hat{Q}_{\pm}=P_{\pm}\gamma^{\mu}\hat{D}_{\mu}\label{eq:18}\end{equation}
with

\begin{equation}
\hat{D}_{\mu}=\hat{p}_{\mu}-eA_{\mu}.\label{eq:19}\end{equation}
These operators are nilpotent:

\begin{equation}
\hat{Q}_{\pm}^{2}=0.\label{eq:20}\end{equation}
With the latter property, one gets

\begin{equation}
\left[\gamma^{\mu}\gamma^{\nu}\hat{D}_{\mu}\hat{D}_{\nu},\,\hat{Q}_{\pm}\right]_{-}=0\label{eq:21a}\end{equation}
and with \eqref{eq:9}, we obtain

\begin{equation}
\left[\hat{Q}_{+},\,\hat{Q}_{-}\right]_{+}=\gamma^{\mu}\gamma^{\nu}\hat{D}_{\mu}\hat{D}_{\nu}.\label{eq:21b}\end{equation}
\eqref{eq:21a} together with \eqref{eq:21b} form a supersymmetric
algebra and the operator $\gamma^{\mu}\gamma^{\nu}\hat{D}_{\mu}\hat{D}_{\nu}$
would be a supersymmetric one, if it were hermitian. Unfortunately,
the latter is only the case, if the Lorentz-metric is replaced by
a euclidian one, i.e.

\begin{equation}
g_{\mu\nu}\rightarrow-\delta_{\mu\nu}.\label{eq:22}\end{equation}

\section*{Lagrangian}

From the subsequent Lagrangian,

\begin{equation}
L_{\pm}=-\left[\left(\hat{p}^{\mu}-M^{\mu T}\right)\bar{\psi}_{\mp}\right]\left[\left(\hat{p}_{\mu}+M_{\mu}\right)\psi_{\pm}\right]-\bar{\psi}_{\mp}m^{2}\psi_{\pm}+\bar{\psi}_{\mp}\frac{1}{2}\sigma^{\mu\nu}E_{\mu\nu}\psi_{\pm},\label{eq:23}\end{equation}
the equations \eqref{eq:14a} and \eqref{eq:15} follow with the aid
of the Euler equations

\begin{equation}
\frac{\partial L_{\pm}}{\partial\bar{\psi}_{\mp}}=\partial_{\mu}\frac{\partial L_{\pm}}{\partial\left(\partial_{\mu}\bar{\psi}_{\mp}\right)}\label{eq:24a}\end{equation}
and

\begin{equation}
\frac{\partial L_{\pm}}{\partial\psi_{\mp}}=\partial_{\mu}\frac{\partial L_{\pm}}{\partial\left(\partial_{\mu}\psi_{\mp}\right)},\label{eq:24b}\end{equation}
respectively.

\section*{The electromagnetic example}

Before we address to the electroweak interaction, we discuss the much
more simple case for an electromagnetic potential, i.e.

\begin{equation}
M_{\mu}=-eA_{\mu}\label{eq:25a}\end{equation}
and

\begin{equation}
E_{\mu\nu}=-eF_{\mu\nu}\label{eq:25b}\end{equation}
with the electromagnetic field tensor

\begin{equation}
F_{\mu\nu}=\partial_{\mu}A_{\nu}-\partial_{\nu}A_{\mu}\label{eq:25c}\end{equation}
as well as $\psi$ from \eqref{eq:1} with only one component and
an electron mass $m$. In this case, the full Lagrangian $L$ is the
sum of $L_{+}$ and $L_{-}$ from \eqref{eq:23}:

\begin{equation}
L=L_{+}+L_{-}\label{eq:26}\end{equation}
because left- and right-handed electrons or positrons should have
equal rights in the pure electromagnetic case. Due to the properties
\eqref{eq:16}, \eqref{eq:8} and \eqref{eq:9}, terms like

\begin{equation}
\bar{\psi}_{\mp}\bar{\psi}_{\pm}=\bar{\psi}P_{\pm}\psi\label{eq:27}\end{equation}
and

\[
\bar{\psi}_{\mp}\sigma^{\mu\nu}\bar{\psi}_{\pm}=\bar{\psi}\sigma^{\mu\nu}P_{\pm}\psi\]
appear in \eqref{eq:26}. But because of

\begin{equation}
\psi=\psi_{+}+\psi_{-},\label{eq:29}\end{equation}
the full Lagrangian becomes just

\begin{equation}
L=-\left[\left(\hat{p}^{\mu}+eA^{\mu}\right)\bar{\psi}\right]\left[\left(\hat{p}_{\mu}-eA_{\mu}\right)\psi\right]-m^{2}\bar{\psi}\psi-\frac{1}{2}eF_{\mu\nu}\bar{\psi}\sigma^{\mu\nu}\psi,\label{eq:30}\end{equation}
which states again, that we do not draw a distinction between left-
and right-handed helicities in this pure electromagnetic case. Therefore
the Euler equations are

\begin{equation}
\frac{\partial L}{\partial\bar{\psi}}=\partial_{\mu}\frac{\partial L}{\partial\left(\partial_{\mu}\bar{\psi}\right)}\label{eq:31a}\end{equation}
and

\begin{equation}
\frac{\partial L}{\partial\psi}=\partial_{\mu}\frac{\partial L}{\partial\left(\partial_{\mu}\psi\right)}.\label{eq:31b}\end{equation}

\section*{The electroweak case}

Let us just cite here the potential for an electroweak coupling (see
e.g. \cite{key-3}):

\begin{equation}
M_{\mu}=\left(\begin{array}{cc}
g\vec{W}_{\mu}\cdot\vec{T}+g^{\prime}Y_{L}B_{\mu} & 0\\
0 & g^{\prime}Y_{R}\, B_{\mu}\end{array}\right),\label{eq:32a}\end{equation}

\begin{equation}
\vec{T}=\frac{1}{2}\vec{\sigma}\label{eq:32b}\end{equation}
with the Pauli-matrices $\vec{\sigma}=\left(\sigma_{1},\sigma_{2},\sigma_{3}\right)$
and the potential

\begin{equation}
\vec{W}_{\mu}=\left(W_{\mu}^{1},W_{\mu}^{2},W_{\mu}^{3}\right)\label{eq:32c}\end{equation}
of a Yang-Mills field with field tensor

\begin{equation}
F_{\mu\nu}=\left(\partial_{\mu}\vec{W}_{\nu}-\partial_{\nu}\vec{W}_{\mu}\right)\cdot\vec{T}+g\left(\vec{W}_{\mu}\times\vec{W}_{\nu}\right)\cdot\vec{T}=\vec{F}_{\mu\nu}\cdot\vec{T}.\label{eq:33}\end{equation}
$\vec{W}_{\mu}$ belongs to the left-handed doublet in \eqref{eq:3},
whereas the potential $B_{\mu}$ of a field 

\begin{equation}
B_{\mu\nu}=\partial_{\mu}B_{\nu}-\partial_{\nu}B_{\mu}\label{eq:34}\end{equation}
is associated with the right-handed singlet in \eqref{eq:3}. $g$
and $g^{\prime}$ are the corresponding coupling constants. The so-called
weak hypercharges $Y_{L}$ and $Y_{R}$ are usually identified with
the values

\begin{equation}
Y_{L}=-\frac{1}{2},\label{eq:35a}\end{equation}

\begin{equation}
Y_{R}=-1.\label{eq:35b}\end{equation}
Instead of the three components $W_{\mu}^{i}$ in \eqref{eq:32c},
one uses mostly

\begin{equation}
\sqrt{2}W_{\mu}^{\pm}=W_{\mu}^{1}\mp iW_{\mu}^{2}\label{eq:36a}\end{equation}
and \begin{equation}
W_{\mu}^{0}=W_{\mu}^{3}.\label{eq:36b}\end{equation}

From the structure of \eqref{eq:32a}, we can see that there is only
a coupling between the left-handed leptons $\nu_{L}$ and $e_{L}$
of the doublet in \eqref{eq:3}, but no coupling between this doublet
and the right-handed singlet $e_{R}$ in \eqref{eq:3}. This is the
reason, why we can construct a common Lagrangian for the left- and
the right-handed doublet in \eqref{eq:3} from \eqref{eq:23}:

\begin{equation}
L=-\left(\hat{p}^{\mu}\bar{\psi}\right)\left(\hat{p}_{\mu}\psi\right)-\bar{\psi}m^{2}\psi-\left(\hat{p}^{\mu}\bar{\psi}\right)\left(M_{\mu}\psi\right)+\left(M^{\mu T}\bar{\psi}\right)\left(\hat{p}_{\mu}\psi\right)+\left(M^{\mu T}\bar{\psi}\right)\left(M_{\mu}\psi\right)+\frac{1}{2}\bar{\psi}\sigma^{\mu\nu}E_{\mu\nu}\psi\label{eq:37a}\end{equation}
with

\begin{equation}
\bar{\psi}=\left(\begin{array}{c}
\bar{\nu}_{R}\\
\bar{e}_{R}\\
\bar{e}_{L}\end{array}\right).\label{eq:37b}\end{equation}
Realize that the helicity of the three 4-spinors forming the components
of $\bar{\psi}$ is the reverse one compared with those in $\psi$
(see equation \eqref{eq:3}). The last term in \eqref{eq:37a} contains
fields:

\begin{equation}
E_{\mu\nu}=\left(\begin{array}{cc}
g\vec{F}_{\mu\nu}\cdot\vec{T}+g^{\prime}Y_{L}\, B_{\mu\nu} & 0\\
0 & g^{\prime}Y_{R}\, B_{\mu\nu}\end{array}\right).\label{eq:37c}\end{equation}
Usually, the case of massless particles is considered, i.e. $m=0$.
From the Lagrangian \eqref{eq:37a}, we can read off a term

\begin{equation}
\left(M^{\mu T}\bar{\psi}\right)\left(M_{\mu}\psi\right)=\bar{\psi}M^{\mu}M_{\mu}\psi.\label{eq:38}\end{equation}
Especially the diagonal terms in $M^{\mu}M_{\mu}$ lead to interesting
results, if we also take into account that the physically relevant
fields \textendash{} apart from $W_{\mu}^{\pm}$ that belong to the
\emph{W} bosons \textendash{} are a combination of $W_{\mu}^{0}$
and $B_{\mu}$,

\begin{equation}
Z_{\mu}=\frac{1}{\sqrt{g^{2}+g^{\prime2}}}\left(gW_{\mu}^{0}-g^{\prime}B_{\mu}\right)\label{eq:39a}\end{equation}
and

\begin{equation}
A_{\mu}=\frac{1}{\sqrt{g^{2}+g^{\prime2}}}\left(g^{\prime}W_{\mu}^{0}+gB_{\mu}\right),\label{eq:39b}\end{equation}
which are known to belong to the \emph{Z} boson and the photon, respectively.
By the way, Lagrangian \eqref{eq:37a} reproduces the Lagrangian \eqref{eq:30}
for a pure electromagnetic interaction, if we set the boson wave operators
$W_{\mu}^{\pm}$ and $Z$ to zero. Of course, this is what we should
have expected from \eqref{eq:37a}. The product $M^{\mu}M_{\mu}$
leads to the subsequent terms in \eqref{eq:38}:

\begin{equation}
M^{\mu}M_{\mu}=\left(\begin{array}{cc}
\left(M^{\prime\mu}M_{\mu}^{\prime}\right)_{2\times2} & 0\\
0 & g^{\prime2}B_{\mu}B^{\mu}\end{array}\right)\label{eq:40}\end{equation}
with\begin{equation}
M^{\prime\mu}M_{\mu}^{\prime}=\left(\begin{array}{cc}
\frac{1}{4}\left(g^{2}+g^{\prime2}\right)Z_{\mu}Z^{\mu}+\frac{1}{2}g^{2}W_{\mu}^{+}W^{-\mu} & a_{12}\\
a_{21} & a_{22}\end{array}\right)\label{eq:40b}\end{equation}

\begin{equation}
a_{12}=\frac{g\,\sqrt{g^{2}+g^{\prime2}}}{2\sqrt{2}}Z_{\mu}W^{+\mu}-\frac{gW_{\mu}^{+}}{2\sqrt{2}\sqrt{g^{2}+g^{\prime2}}}\left[\left(g^{2}-g^{\prime2}\right)Z^{^{\mu}}+2gg^{\prime}A^{\mu}\right],\label{eq:40c}\end{equation}

\begin{equation}
a_{21}=\frac{g\,\sqrt{g^{2}+g^{\prime2}}}{2\sqrt{2}}Z_{\mu}W^{-\mu}-\frac{gW_{\mu}^{-}}{2\sqrt{2}\sqrt{g^{2}+g^{\prime2}}}\left[\left(g^{2}-g^{\prime2}\right)Z^{^{\mu}}+2gg^{\prime}A^{\mu}\right],\label{eq:40d}\end{equation}

\begin{equation}
a_{22}=\frac{1}{2}g^{2}W_{\mu}^{+}W^{-\mu}+\frac{1}{4\left(g^{2}+g^{\prime2}\right)}\left[\left(g^{2}-g^{\prime2}\right)Z^{^{\mu}}+2gg^{\prime}A^{\mu}\right]^{2}.\label{eq:40e}\end{equation}

The most remarkable term in $M^{\prime\mu}M_{\mu}^{\prime}$ of \eqref{eq:40}
is its first element, since it resembles the mass terms for the \emph{Z}
and \emph{W} bosons arising from the presence of a Higgs field. Usually,
the latter field can be described by two complex bosonic fields $\phi_{1}$
and $\phi_{2}$ forming an SU(2)-doublet:

\begin{equation}
\phi\left(x\right)=\left(\begin{array}{c}
\phi_{1}\\
\phi_{2}\end{array}\right).\label{eq:41}\end{equation}
The procedure of spontaneous symmetry breaking gives a field e.g.
of the form (see e.g. \cite{key-3})

\begin{equation}
\phi\left(x\right)=\frac{1}{\sqrt{2}}\left(\begin{array}{c}
0\\
u+H\left(x\right)\end{array}\right)\label{eq:42}\end{equation}
with a constant $u$. Since $\phi$ is bosonic, it obeys a Klein-Gordon-like
equation that contains a kinetic term

\begin{equation}
\hat{D}_{\mu}\phi^{\dagger}\hat{D}^{\mu}\phi,\label{eq:43}\end{equation}
where $\hat{D}_{\mu}$ denotes a covariant derivative

\begin{equation}
\hat{D}_{\mu}=\partial_{\mu}+ig\vec{W}_{\mu}\cdot\vec{T}+ig^{\prime}Y_{\phi}B_{\mu}\label{eq:44}\end{equation}
with a weak hypercharge of the field $\phi$,

\begin{equation}
Y_{\phi}=\frac{1}{2}.\label{eq:45}\end{equation}
Together with \eqref{eq:42}, \eqref{eq:44} and \eqref{eq:45}, the
kinetic term \eqref{eq:43} produces a term

\begin{equation}
\left[\frac{1}{8}\left(g^{2}+g^{\prime2}\right)Z_{\mu}Z^{\mu}+\frac{1}{4}g^{2}W_{\mu}^{+}W^{-\mu}\right]u^{2}=\frac{1}{2}m_{Z}^{2}\, Z_{\mu}Z^{\mu}+m_{W}^{2}\, W_{\mu}^{+}W^{-\mu}\label{eq:46}\end{equation}
with the \emph{W} and \emph{Z} boson masses

\begin{equation}
m_{W}=\frac{1}{2}gu\label{eq:47a}\end{equation}
and

\begin{equation}
m_{Z}=\frac{1}{2}\sqrt{g^{2}+g^{\prime2}}\, u,\label{eq:47b}\end{equation}
respectively. This might lead us to the speculation that with the
Lagrangian \eqref{eq:37a}, we can repeat this procedure of spontaneous
symmetry breaking with neutrino fields instead of the Higgs boson
fields. That means, we would like to perform replacements like the
following ones, analogous to \eqref{eq:42}:

\begin{equation}
\psi=\left(\begin{array}{c}
\nu_{L}\\
e_{L}\\
e_{R}\end{array}\right)\rightarrow\left(\begin{array}{c}
u_{L}+H_{L}\left(x\right)\\
0\\
0\end{array}\right),\;\bar{\psi}=\left(\begin{array}{c}
\bar{\nu}_{R}\\
\bar{e}_{R}\\
\bar{e}_{L}\end{array}\right)\rightarrow\left(\begin{array}{c}
\bar{u}_{R}+\bar{H}_{R}\left(x\right)\\
0\\
0\end{array}\right)\label{eq:48}\end{equation}
(where we have omitted a factor similar to the $\frac{1}{\sqrt{2}}$
in \eqref{eq:42}). If we do this, only the term with \begin{equation}
\frac{1}{4}\left(g^{2}+g^{\prime2}\right)Z_{\mu}Z^{\mu}+\frac{1}{2}g^{2}W_{\mu}^{+}W^{-\mu}\label{eq:49}\end{equation}
remains non-zero in $\bar{\psi}M^{\mu}M_{\mu}\psi$. If we consider
$u_{L}$ and $\bar{u}_{R}$ to be constants, their derivatives would
vanish, as it was the case for the bosonic $u$ of the Higgs field,
but the non-kinetic term $\left(M^{\mu T}\bar{\psi}\right)\left(M_{\mu}\psi\right)+\frac{1}{2}\bar{\psi}\sigma^{\mu\nu}E_{\mu\nu}\psi$
in \eqref{eq:37a} does not disappear. One of the possible objections
to this procedure is that this way, we would get rid of terms in the
Lagrangian \eqref{eq:37a}, which we normally need for the description
of electrons and neutrinos. And of course, this was really not our
intention! However, realize that we could construct a counter-part
of the Lagrangian \eqref{eq:37a} with reversed helicities of the
neutrino and electron spinors, i.e. with \begin{equation}
\psi=\left(\begin{array}{c}
\nu_{R}\\
e_{R}\\
e_{L}\end{array}\right)\rightarrow\left(\begin{array}{c}
u_{R}+H_{R}\left(x\right)\\
0\\
0\end{array}\right),\;\bar{\psi}=\left(\begin{array}{c}
\bar{\nu}_{L}\\
\bar{e}_{L}\\
\bar{e}_{R}\end{array}\right)\rightarrow\left(\begin{array}{c}
\bar{u}_{L}+\bar{H}_{L}\left(x\right)\\
0\\
0\end{array}\right).\label{eq:50}\end{equation}
In this case, we would be glad to be able to eliminate the neutrino
and anti-neutrino terms with a helicity that is not observable in
nature. We would also be happy to be able to eliminate the electron
terms in that Lagrangian, because our original Lagrangian \eqref{eq:37a}
already considers everything we need for the description of electrons.
Thus our final speculation is that there is only one sort of neutrinos
with respect to their helicity, because the second sort (i.e. the
right-handed one not present in nature) is possibly \textquotedblleft{}needed\textquotedblright{}
to provide the \emph{W }and \emph{Z} gauge bosons of the weak interaction
with their masses. 

On the other hand, it is not obvious, what a term like $\frac{1}{2}\bar{\psi}\sigma^{\mu\nu}E_{\mu\nu}\psi$
will produce within the above-sketched procedure of symmetry breaking
with neutrinos (instead of Higgs bosons). 

In order to verify (or falsify) these ideas, we would have to show
that the Lagrangian \eqref{eq:37a} is (or is not) able to reproduce
the results of the Standard Model with respect to the electro-weak
interaction. However, e.g. the structure of the term \eqref{eq:40b}
makes clear that this might not be an easy task. Hence, one of the
first steps should rather be to investigate, whether the Lagrangian
\eqref{eq:30} for a pure electromagnetic interaction can reproduce
the well-known results of Quantum Electrodynamics.

\section*{Conclusions}

We have seen that the first element in the matrix of \eqref{eq:40b}
\textendash{} i.e. the element only connected with neutrino fields
\textendash{} resembles the term \eqref{eq:46} with a Higgs boson
field. The latter is known to produce the masses of the \emph{W} and
\emph{Z} gauge bosons of the weak interaction within the Standard
Model. This invites us to the speculation that neutrinos and Higgs
bosons are relatives, although the former are spin-1/2 and the latter
spin-0 particles. But the connection of Lagrangians of the kind of
\eqref{eq:23} with chiral supersymmetry might even be a hint that
the relation between both sorts of particles could be a supersymmetric
one. We have also formulated the idea that the missing right-handed
sort of neutrinos in nature could have assumed the role \textendash{}
normally ascribed to the Higgs bosons \textendash{} to provide the
\emph{W} and \emph{Z} gauge bosons of the weak interaction with their
masses. This might be a \textquotedblleft{}reason\textquotedblright{}
why nobody has found right-handed neutrinos in experiments.

\end{document}